# The Structural, Thermodynamic and Electronic Properties of $Zr_{0.75}Hf_{0.25}Co$ and $Zr_{0.75}Ti_{0.25}Co$：A First Principle Study

Gan Ren [*]

Departments of Physics & Key laboratory of photonic and optical detection in civil aviation, Civil Aviation Flight University of China, Guanghan 618307, China

**Abstract**

ZrCo alloy is promising to substitute uranium for handling hydrogen isotope storage in thermonuclear reactor. The alloying substitution of Zr in ZrCo with Hf or Ti can enhance the ability of anti-disproportionation. In this work, $Zr_{0.75}Hf_{0.25}Co$ and $Zr_{0.75}Ti_{0.25}Co$ were considered in the framework of density functional theory, aimed to investigate the properties of the alloying substitution of Zr with Hf or Ti in ZrCo. Our results found that the optimized lattice constants of the alloying substitutions, $Zr_{0.75}Hf_{0.25}Co$ and $Zr_{0.75}Ti_{0.25}Co$, are smaller than ZrCo. The thermodynamic stability reduces in the order $Zr_{0.75}Hf_{0.25}Co > ZrCo > Zr_{0.75}Ti_{0.25}Co$, as demonstrated by the enthalpy of formation. The valence electrons are mainly localized at the ion core and the chemical bonds are polarized in $Zr_{0.75}Hf_{0.25}Co$ and $Zr_{0.75}Ti_{0.25}Co$ analogous to ZrCo.

**Keywords**

ZrCo alloy, alloying substitution, Density functional theory, Electronic structure, Enthalpy of formation

## 1. Introduction

Uranium alloy is applied to handle hydrogen isotopes storage in the International Thermonuclear Experimental Reactor (ITER) system for its favorable tritium storage properties [1-3]. However, some disadvantages, such as high pyrophoricity, pulverization and nuclear materials, hinder its further application [4, 5]. Due to the favorable properties for handling recovery, storage and supply of tritium, ZrCo intermetallic was proposed to be a promising substitute for tritium handling in ITER. ZrCo alloy exhibits comparable favorable properties for handling tritium like uranium alloy [6-9], such as low equilibrium adsorption pressure ($10^{-3}$Pa) at room temperature and high hydrogen capacity per unit mass (1.96 wt %). Besides, ZrCo alloy is more stable and safer than uranium alloy for its low pyrophoricity and non-ridioactivation.

Nonetheless, the disproportionation of ZrCo hydride (ZrCoH$_x$ ($x\leq3$)) blocks the application of ZrCo alloy in practice; which significantly degrades the capacity of hydrogen storage and blocks its recovery. In disproportionation process, ZrCoH$_x$ decomposes into the thermodynamically more stable ZrH$_2$ and ZrCo$_2$ [10, 11], but not ZrCo and H$_2$. The disproportionation follows the reaction [12]

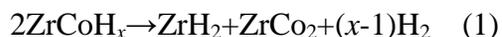
$$2ZrCoH_x \rightarrow ZrH_2 + ZrCo_2 + (x-1)H_2 \quad (1)$$

The hydride ZrH$_2$ requires much higher temperature about 973 K to release H$_2$. ZrCo$_2$ doesn't adsorb hydrogen and is not active for recovery.

To decrease the disproportionation, many studies were conducted by alloying substitution to improve the thermal stability of ZrCo and reduce the tendency of hydrogen-induced disproportionation. Alloying substitution of Zr with Hf or Ti in ZrCo has gained much attention for similar valence electron distribution among Zr, Hf and Ti. Zr$_{1-x}$Ti$_x$Co ($x$ = 0, 0.1, 0.2, 0.3) [13, 14] and Zr$_{1-x}$Hf$_x$Co ($x$ = 0, 0.1, 0.2, 0.3) [15, 16] alloys were found to easily form a solid solution; both exhibit a single flat plateau in isothermal hydrogen sorption process. Huang et al. [13] reported that Zr$_{1-x}$Ti$_x$Co ($x$ = 0, 0.1, 0.2, 0.3) alloys is also in a cubic crystal structure analogous to ZrCo. The hydrogen equilibrium pressure increases with increasing addition of Ti, but the desorption temperature for supplying 100 kPa decreases with the addition of Ti. Konishi et al. [15] reported Zr$_{1-x}$Hf$_x$Co (x<0.5) alloys are also in a single cubic phase. A single sloping plateau was observed for all isotherms. The equilibrium pressure increases with increasing Hf contents at the same temperature analogous to the substitution with Ti. Moreover, Zr$_{1-x}$Ti$_x$Co and Zr$_{1-x}$Hf$_x$Co have an enhanced ability of anti-disproportionation compared with ZrCo [15, 17, 18].

Very few theoretical studies have been reported on the ZrCo systems especially its alloying substitution [19-24]. Li et al. [20] investigated the structural, vibrational and thermodynamic properties of ZrCo by first principles. They calculated the phonon dispersion curves of ZrCo via density functional perturbation theory and further evaluated the thermodynamics. Gupta [19] calculated the electronic structure of $ZrCoH_3$ via ab-initio self-consistent linear muffin-tin orbital method. He found that Zr-H bond is essential to the stability of the $ZrCoH_3$ for lowering of the energy states below the Fermi energy. Chattaraj et al. [21-23] explored the structural, thermodynamic, electronic and phonon dispersion properties of ZrCo, $ZrCoX_3$ (X = H, D and T) and $ZrX_2$ (X = H, D and T) in terms of the density functional theory approach. They found $ZrCoH_3$ is thermodynamically more stable than ZrCo; chemical bonds are polarized in ZrCo and $ZrCoH_3$, besides the valence electrons are mainly localized at the ion core [21]. The energy gap between the optical and acoustic modes is reduced in the order $ZrCoT_3 > ZrCoD_3 > ZrCoH_3$. ZrCo intermetallic is dynamically stable whereas $ZrCoX_3$ (X = H, D and T) are dynamically unstable. The thermodynamic stability of $ZrCoX_3$ (X = H, D and T) reduces in the order $ZrCoT_3 > ZrCoD_3 > ZrCoH_3$. ZrCo and $ZrCoH_3$ are mechanically stable by analyzing elastic properties [22]. The phonon energy gap formation energies between optical and acoustic modes reduce more for $ZrT_2$ than $ZrD_2$ and $ZrH_2$. The formation energies of $ZrX_2$ follows the order $ZrT_2 < ZrD_2 < ZrH_2$ [23]. Yang et al. [24] investigated the alloy substitution of Zr or Co with Ti, Hf, Sc, Fe, Ni, Cu as well as the substitution influences on hydrogen-induced disproportionation of ZrCo alloy. They proposed that H at 8e site of $ZrCoH_3$ (H(8e)) plays the key role in the disproportionation process; H(8e) prefers to form strong covalent-like binding with neighboring Co as well as its substitute elements. To the best of our knowledge so far, no reports are available to describe the electronic structure and chemical bonding of $Zr_{1-x}Ti_xCo$ and $Zr_{1-x}Hf_xCo$ using the first principle theory. In this work, $Zr_{0.75}Hf_{0.25}Co$ and $Zr_{0.75}Ti_{0.25}Co$ were adopted to investigate the structure, thermodynamic and electronic properties of the alloying substitution of Zr with Hf or Ti in ZrCo in terms of the density functional theory.

2. Computational details

All the calculations were carried out with the Vienna ab-initio simulation package (VASP) [25-27] in terms of spin-polarized density functional theory (DFT). The electron-ion interaction was described by the projector-augmented wave (PAW) method [28, 29]. The wave functions were expanded in a plane-wave basis set with a energy cutoff of 500 eV. The exchange correlation energy was described by generalized gradient approximation (GGA) of Perdew-Burke-Ernzerhof (PBE) [30] scheme. The conjugate gradient method was adopted to perform ionic optimization and the forces on each ion were minimized to 5meV/Å [21, 31]. The convergence criteria for energy were $10^{-5}$ eV.

It has been found that $Zr_{1-x}Ti_xCo$ ($x$ = 0, 0.1, 0.2, 0.3) [13, 14] and $Zr_{1-x}Hf_xCo$ ($x$ = 0, 0.1, 0.2, 0.3) [15, 16] is also cubic structure like ZrCo. For simplicity, we adopted $Zr_{0.75}Hf_{0.25}Co$ and $Zr_{0.75}Ti_{0.25}Co$ with the crystal structure plotted in Figure 1 to evaluate the properties of alloying substitution of Zr with Hf and Ti in ZrCo.

## 3. Results and discussion

The equilibrium structures were first determined by evaluating the total energy (E) with different unit cell volumes (V). The obtained E-V data was then fitted using the Murnaghan equation of state [32] to determine the equilibrium volume and bulk moduli at temperature T = 0K. The lattice parameters of $α$-Zr, Co, Hf, Ti, ZrCo, $Zr_{0.75}Ti_{0.25}Co$ and $Zr_{0.75}Hf_{0.25}Co$ were obtained with the procedures. All optimized lattice parameters are collected in Table 1.

The crystal structure of ZrCo is CsCl-type cubic (bcc) and the lattice constant $a$ = 3.196 Å [33]. The optimized $a$ = 3.182 Å for ZrCo. The optimized lattice parameters of $α$-Zr, Co, Hf, Ti and ZrCo have a good agreement with the experimental data that the deviations are within ±1%. The atomic and electronic structure of $Zr_{0.75}Hf_{0.25}Co$ and $Zr_{0.75}Ti_{0.25}Co$ were optimized, the energy varied with volume curves (E-V) were plotted in Figure 2. The equilibrium volume and bulk moduli were fitted by Murnaghan equation of state [32] and listed in Figure 2. Both lattice parameters of $Zr_{0.75}Ti_{0.25}Co$ and $Zr_{0.75}Hf_{0.25}Co$ are smaller than ZrCo. The lattice parameters increase in the order $Zr_{0.75}Ti_{0.25}Co$ <$Zr_{0.75}Hf_{0.25}Co$<ZrCo. The results are agreement with previous experiment results [14, 18]; the lattice constant of ZrCo decreases with the substitution of Zr with Ti or Hf in ZrCo. TiCo and HfCo are also CsCl-type cubic (bcc) and with lattice constant smaller than ZrCo. Because $Zr_{1-x}Ti_xCo$ ($x$ = 0, 0.1, 0.2, 0.3) [13, 14] and $Zr_{1-x}Hf_xCo$ ($x$ = 0, 0.1, 0.2, 0.3) [15, 16] alloys form a solid solution, the lattice constant of ZrCo decreases after addition Ti or Hf as proposed by Vegard rule.

After get the equilibrium crystal structure, enthalpies of formation for ZrCo, $Zr_{0.75}Hf_{0.25}Co$ and $Zr_{0.75}Ti_{0.25}Co$ were calculated to evaluate their thermodynamic stability. The enthalpy of formation [21] at T=0K was defined as

$$\triangle_f H(Zr_{1-x}X_xCo) = E_{tot}(Zr_{1-x}X_xCo) - (1-x)E_{tot}(Zr) - x E_{tot}(X) - E_{tot}(Co) \quad (2)$$

where X is Zr, Hf or Ti; $x = 0$ for Zr and 0.25 for Hf or Ti, respectively. The total energies of each compound were calculated and collected in Table 2. The enthalpies of formation for ZrCo, $Zr_{0.75}Hf_{0.25}Co$ and $Zr_{0.75}Ti_{0.25}Co$ at T=0K were estimated to be -53.5kJ/mol, -58.6kJ/mol and -49.5kJ/mol, respectively. The thermodynamic stability of ZrCo, $Zr_{0.75}Hf_{0.25}Co$ and $Zr_{0.75}Ti_{0.25}Co$ reduces in the order $Zr_{0.75}Hf_{0.25}Co>ZrCo>Zr_{0.75}Ti_{0.25}Co$. The results suggest that substitution of Zr with Hf in ZrCo improve the thermodynamic stability of ZrCo alloy [15, 16], but substitution of Zr with Ti doesn't. The lattice parameter of $Zr_{0.75}Ti_{0.25}Co$ and $Zr_{0.75}Hf_{0.25}Co$ are smaller than ZrCo, the improved anti-disproportionation properties may be mainly introduced by the decrease of lattice parameter as previous experiments proposed [18, 24], which need further research.

To understand the chemical bonding of $Zr_{0.75}Ti_{0.25}Co$ and $Zr_{0.75}Hf_{0.25}Co$, the total density of states spectrum of $Zr_{0.75}Ti_{0.25}Co$ and $Zr_{0.75}Hf_{0.25}Co$ as well as its constituent elements were calculated and were plotted in Figure 3 and Figure 4, respectively. The d-orbital electrons are mainly contributing at the Fermi level for Zr, Co, Hf and Ti, analogous to Zr and Co in ZrCo [21]. Because Zr, Hf and Ti are in the same main group and the valence electrons are the same, the electron densities of states are almost the same for Zr, Hf and Ti. Besides the d-orbital spectrum of Co is more localized and contributing more than the d-orbital spectrum of Zr, Hf and Ti. The d-orbital spectrum of Zr, Co Hf and Ti are overlapped around at the Fermi energy which ensures the metallic nature of $Zr_{0.75}Ti_{0.25}Co$ and $Zr_{0.75}Hf_{0.25}Co$. Similar phenomena have also been observed in ZrCo [21]. Overall, alloying substitution of Zr with Hf or Ti in ZrCo has little influence on the electron density of states of ZrCo for the same valence electrons distribution for Zr, Hf and Ti.

To further understanding the chemical bonding in $Zr_{0.75}Ti_{0.25}Co$ and $Zr_{0.75}Hf_{0.25}Co$, the deformation charge density distribution characterized the electron transfer before and after bond formation in $Zr_{0.75}Ti_{0.25}Co$ and $Zr_{0.75}Hf_{0.25}Co$ has been calculated. The deformation charge densities along $(\bar{1}10)$ plane are plotted in Figure 5. Chattaraj et al. [21] reported the bond Zr-Co in ZrCo is polarized and electrons are mainly localized at ion core. Fig. 5 shows the maximum of electron transfer from the ion core to the intermediate is about 0.37. The electron transfer of Co is more than Zr, Hf and Ti. The electron transfer is almost the same for Zr, Hf and Ti. The electron transfer is small compared with valence electrons of each ion. The results suggest that electrons are also mainly localized at ion core analogous to that in ZrCo. Moreover, because Co has more valence electrons than Zr, Hf and Ti; the Zr-Co, Hf-Co and Ti-Co bonds in $Zr_{0.75}Ti_{0.25}Co$ and $Zr_{0.75}Hf_{0.25}Co$ are also polarized.

## 4. Conclusions

In summary, we have calculated the structural, thermodynamic and electronic properties of $Zr_{0.75}Hf_{0.25}Co$ and $Zr_{0.75}Ti_{0.25}Co$ in terms of spin-polarized density functional theory approach. The optimized lattice parameters are good agreement with the experimental data. The lattice parameters of $Zr_{0.75}Hf_{0.25}Co$ and $Zr_{0.75}Ti_{0.25}Co$ are smaller than ZrCo. The lattice parameters reduce in the order ZrCo> $Zr_{0.75}Hf_{0.25}Co>Zr_{0.75}Ti_{0.25}Co$. The thermodynamically stability reduces in the order $Zr_{0.75}Hf_{0.25}Co>$ ZrCo> $Zr_{0.75}Ti_{0.25}Co$ as shown by the calculated formation enthalpies. Because Zr, Hf and Ti are in the same group, no explicitly differences have been observed in the electron density of states among ZrCo, $Zr_{0.75}Hf_{0.25}Co$ and $Zr_{0.75}Ti_{0.25}Co$. The electrons in $Zr_{0.75}Hf_{0.25}Co$ and $Zr_{0.75}Ti_{0.25}Co$ are localized at ion core and the bonds polarized analogous to that observed in ZrCo. Overall, alloying substitution of Zr with Hf or Ti in ZrCo mainly influences the lattice constant and thermodynamic stability of ZrCo, but weakly influence on the electronic structure for the same valence electrons.

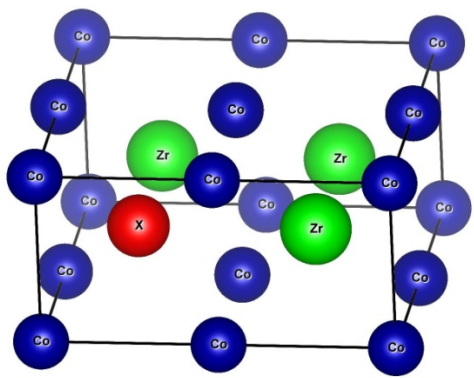

**Figure 1.** *The crystal structure of $Zr_{0.75}Hf_{0.25}Co$ and $Zr_{0.75}Ti_{0.25}Co$ adopted in this work, ion X is Hf or Ti.*

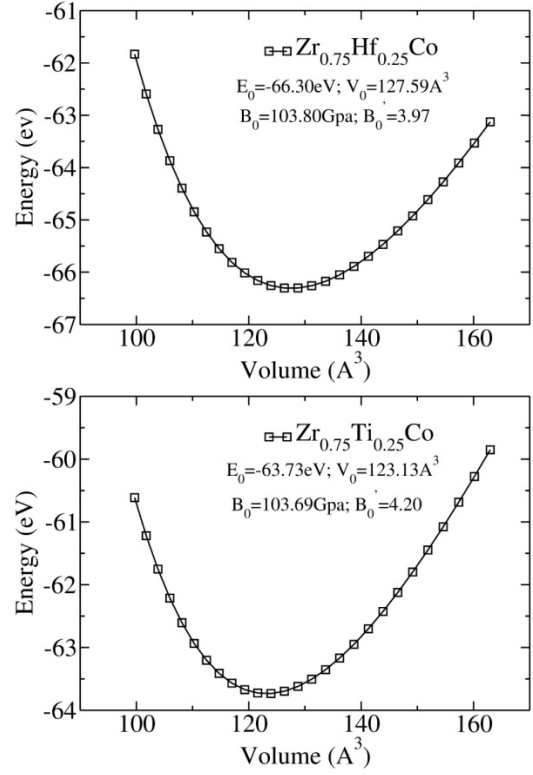

**Figure 2.** *The total energy varies with crystal cell volume of $Zr_{0.75}Hf_{0.25}Co$ and $Zr_{0.75}Ti_{0.25}Co$.*

**Table 1.** *Crystal structure data of α-Zr, Co, Hf, Ti, ZrCo, $Zr_{0.75}Hf_{0.25}Co$ and $Zr_{0.75}Ti_{0.25}Co$*

| System | Crystal system | Space group | Calc. lattice parameters (0 K) Å and volume Å$^3$ | Expt.(298K) Å |
|---|---|---|---|---|
| α-Zr | hcp | P63/mmc | a = 3.234 | 3.233 [34] |
| | | | c = 5.173 | 5.150 |
| | | | $V_0$ = 46.86 | 46.70 |
| Co | hcp | P63/mmc | a = 2.484 | 2.507 [35] |
| | | | c = 4.039 | 4.070 |
| | | | $V_0$ = 21.59 | 22.15 |
| Hf | hcp | P63/mmc | a = 3.194 | 3.193 [34] |
| | | | c = 5.043 | 5.052 |
| | | | $V_0$ = 44.55 | 44.61 |
| Ti | hcp | P63/mmc | a = 2.923 | 2.950 [36] |
| | | | c = 4.625 | 4.681 |

| | | | | | |
|---|---|---|---|---|---|
| ZrCo | bcc | Pm-3m | $V_0 = 34.23$ | 35.30 | |
| | | | $a = 3.182$ | 3.196 [33] | |
| Zr$_{0.75}$Hf$_{0.25}$Co | bcc | Pm-3m | $V_0 = 32.12$ | 32.65 | |
| | | | $a = 3.171$ | | |
| Zr$_{0.75}$Ti$_{0.25}$Co | bcc | Pm-3m | $V_0 = 31.90$ | | |
| | | | $a = 3.134$ | | |
| | | | $V_0 = 30.78$ | | |

***Table 2.*** *Total energies ($E_{tot}$) of α-Zr, Co, Hf, Ti, ZrCo, Zr$_{0.75}$Hf$_{0.25}$Co, Zr$_{0.75}$Ti$_{0.25}$Co and enthalpy formation of ZrCo, Zr$_{0.75}$Hf$_{0.25}$Co, Zr$_{0.75}$Ti$_{0.25}$Co*

| system | $E_{tot}$ (eV)/f.u. Ionic opt | system | $\triangle_f H(0K)$ (kJ/mol) (this work) | $\triangle_f H(298K)$ (experimental) |
|---|---|---|---|---|
| α-Zr | -8.48 | ZrCo | -53.5 | -42.2±1.0 [37] |
| Co | -7.12 | Zr$_{0.75}$Hf$_{0.25}$Co | -58.6 | |
| Hf | -9.96 | Zr$_{0.75}$Ti$_{0.25}$Co | -49.5 | |
| Ti | -7.76 | | | |
| ZrCo | -16.15 | | | |
| Zr$_{0.75}$Hf$_{0.25}$Co | -16.58 | | | |
| Zr$_{0.75}$Ti$_{0.25}$Co | -15.93 | | | |

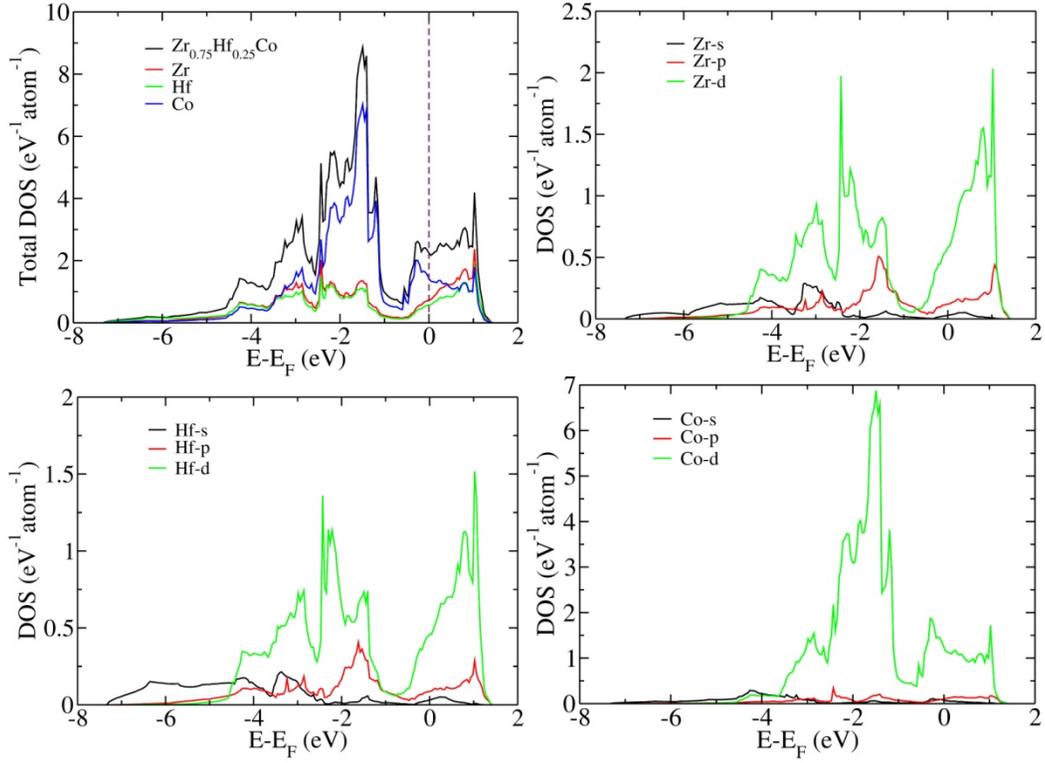

*Figure 3.* *The total and orbital projected density of states (DOS) of Zr$_{0.75}$Hf$_{0.25}$Co.*

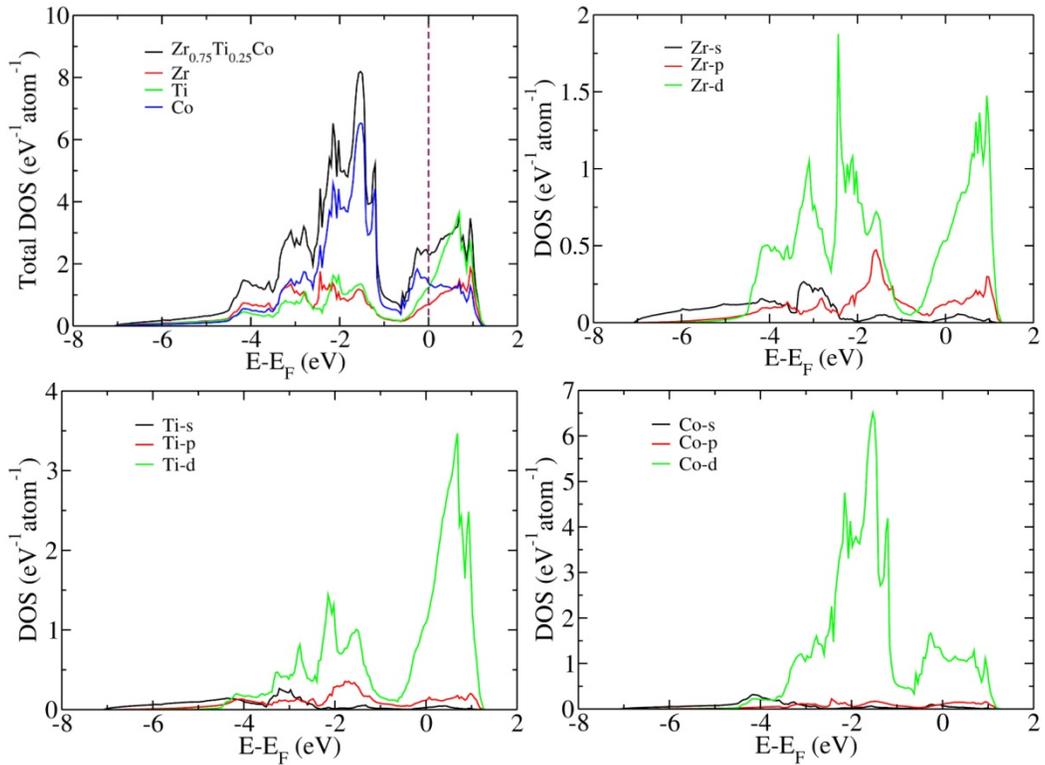

**Figure 4.** *The total and orbital projected density of states (DOS) of $Zr_{0.75}Ti_{0.25}Co$.*

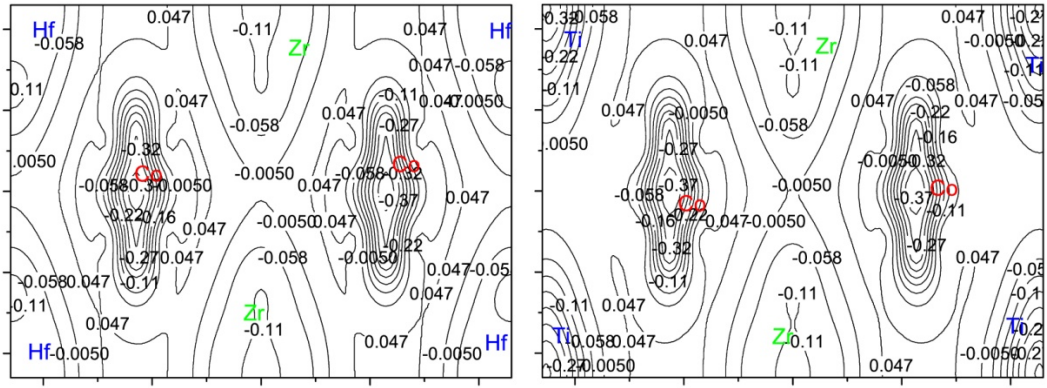

**Figure 5.** *The deformation charge density contour of $Zr_{0.75}Hf_{0.25}Co$ and $Zr_{0.75}Ti_{0.25}Co$ alloy along ($\bar{1}10$) plane.*

**Acknowledgements**

The author thanks Tianhe-2 for allocations of computer time.